# Distributed Agile Software Development:
# A Review

Suprika Vasudeva Shrivastava and Hema Date

**Abstract---** Distribution of software development is becoming more and more common in order to save the production cost and reduce the time to market. Large geographical distance, different time zones and cultural differences in distributed software development (DSD) leads to weak communication which adversely affects the project. Using agile practices for distributed development is also gaining momentum in various organizations to increase the quality and performance of the project. This paper explores the intersection of these two significant trends for software development i.e. DSD and agile. We discuss the challenges faced by geographically distributed agile teams and proven practices to address these issues, which will help in building a successful distributed team.

**Index Terms---** Distributed software development (DSD), global software development (GSD), agile practices, distributed agile development

———————————— ◆ ————————————

## 1  INTRODUCTION

IN the last decade, a great investment is being made to convert national market to global market.

This scenario involves more competition and collaboration [1]. Various challenges like more faults in the projects and scarcity of resources have to be dealt with.

Thus many organizations began to experiment with Distributed Software Development (DSD) facilities to solve these problems. This helped in reducing the cost involved and access to skilled resources [2]. Their main objective is to develop high quality products at lower cost than co-located developments by optimizing the resources [14]. Sometimes, the search for competitive advantage forces organizations to search for external solutions in other countries, which is called Global Software Development (GSD) [3].

Software is developed in a multi-site, multicultural, globally distributed environment. Engineers, managers and executives face formidable changes on many levels, from technical to social and cultural [1]. It affects the way the software is designed, implemented delivered to the customer.

Moreover, since past few years software development community is using agile methods for software development. Agile software development refers to a group of software development methodologies based on iterative development, where requirements and solutions evolve through collaboration between self-organizing cross-functional teams [4]. Agile methods works very well in highly dynamic business and IT environment. Many organizations that develop software using Agile have started looking for skills and talent available at much lower rates and are anxious to source the development work to these centers [5].

This paper aims at understanding the benefits achieved by combining agile with DSD. What are the challenges which agile principles help overcome in projects which are geographically distributed? Are there any new challenges which agile distributed teams have to face, if so, what are the techniques which can be used to handle them?

The paper is structured as follows: Section 2 contains a discussion on Distributed software development (DSD) / Global software development (GSD) and the challenges faced by the geographically distributed team. Section 3 gives an overview of agile methodology. This is followed by section 4, which discusses in details the benefits and challenges faced by distributed agile teams. Section 5 lists down various techniques which can help to overcome the challenges put by agile techniques when combined with distributed teams and last section concludes the paper.

————————————————

- *Mrs. Suprika Vasudeva Shrivastava is with the Symbiosis Centre for Information Technology,P-15 Rajiv Gandhi Infotech Park, Hinjawadi, Pune, India.*
- *Dr.. (Mrs.) Hema Date is with the Department of Information Technology, National Institute of Industrial Engineering (NITIE), Mumbai, India.*

# 2 DISTRIBUTED SOFTWARE DEVELOPMENT

Software has become a core component of business. Success depends on using software to have an edge over the other competitors. Many organizations have started developing software remotely in order to achieve lower cost and access to skilled resources. Moreover large investments have enabled a move from local to global markets in the process of creating new competition and collaboration forms [8].

Various factors are responsible for such a situation:

- The business market proximity advantages including knowledge of customers and local conditions.
- Pressure to improve time-to-market by using time-zone differences and having round-the clock development.
- Create a pool of globally available skilled resources to develop software in reduced cost [1].
- Distribution also minimizes the risk in case of natural catastrophes and other events [9].

Thus software development is now becoming muti-site, multicultural, globally distributed undertaking. More number of organizations are distributing their software development process worldwide to achieve higher profits, productivity, quality and lower cost. This change is having a profound impact not only on marketing and distribution but also on the way products are conceived, designed, constructed, tested and delivered to customers [3].

There are four different ways to distribute software development. Distribution can be defined by its geographical location and by the control and ownership structure of the project. The control structure can be defined by two dimensions: outsourcing means that the company buys the software from some external company and insourcing means that the company provides the services itself through some internal projects. Geographical location is defined by dimensions: onshore, which means that all company's development work takes place in the same country where the headquarters and other operations are located and offshore means that part of the development happens abroad. Onshore distribution is also called distributed software development or DSD and offshore distribution is called global software development, GSD [9].

Thus as a part of globalization effort in the society, software project teams have become globally distributed on the world-wide scale. This characterizes GSD. Various features which distinguishes Global software Development from normal (centralized): distance (distance of developers from each other and from their customers or end users); time-zone differences (time zone is to a large extent a confounding factor with distance); and national culture (including language, national traditions, customs, and norms of behavior) [10].

Along with the benefits obtained through globally distributed development, there are many difficulties faced by various organizations. These problems are caused mainly by distance, time, and cultural differences [13].

Various problems encountered are as follows:

**Strategic issues:** A decision needs to be taken whether the project should be developed by globally dispersed teams or by co-locate teams. Moreover how to divide the work across various sites is also required. Some risk and benefit analysis needs to be done at this stage. Moreover solutions are constrained by the resources available at various sites, their level of expertise in various technologies, infrastructure, etc. Various models are possible and appropriate under different circumstances [1].

**Cultural issues:** GSD requires close cooperation of individuals with different cultural backgrounds. Culture differs on many critical dimensions such as national, professional, ethic, organizational, professional, technical, and team culture [10]. Cultural differences often exacerbate communication problems [12].

**Inadequate communication:** software development requires great deal of formal communication through vital communication channels. The complex infrastructure for GSD and great size of personnel network which changes over time, lead to decrease in communication frequency and quality, which directly affects productivity [14]. Time zone differences of the distributed team further add to the problem.

**Knowledge management:** the team members' experience, methods, decisions and skills must be accumulated during the development process through the effective information sharing so that the team members can use the experience of his predecessor hence helping in reducing cost and redundant work. Without effective information and knowledge-sharing mechanism, managers cannot exploit he GSD's benefits. Distributed environments must facilitate knowledge sharing by maintaining a product/process repository where the contents from various sources like e-mail, online discussions can be kept.

**Project and process management issues:** high organizational complexity, scheduling, task assignment, and cost estimation becomes more problematic in distribute environments as a result of volatile requirements, changing specifications, cultural diversity, and lack of informal communication [15]. Managers must control the overall development process, improving it during the enactment and minimizing any factors that may decrease productivity, taking into account the possible impact of diverse cultures, identifying interrelated tasks, and minimizing dependencies among distributed groups [14].

**Technical issues:** When teams are working across sites, the lack of synchronization can be particularly critical. There is a need to assure commonly defined milestones and clear entry and exit criteria for all tasks. Since networks spanning globally dispersed locations are often slow and unreliable, tasks such as configuration management that involve transmission of critical data must be meticulously planned and executed.

**Risk Management:** risk management is a critical project management activity. DSD development includes issues related to coordination, problem resolution, evolving requirements, and knowledge sharing and risk identification [11]. Software defects become more frequent due to the added complexity, and in most cases, this is related to communication problems and a lack of group awareness. Defects control must be adapted by making a greater effort in risk management activities. The use of adequate measures and the requirements definition is important key factors [14].

As regular face to face communication is becoming very expensive between globally distributed development teams numerous ways have been proposed for the solution to ease the communication challenges in GSD. This includes tools, practices, special persons and reasonable modularization of production across sites. Agile principles have also been suggested as a solution to communication challenges related to DSD [17]. Further the pros and cons of combining agile with DSD will be discussed.

## 3   AGILE PRACTICES

Before going ahead with distributed agile software development, we will have a brief overview some popular agile practices.

Agile software development refers to a group of software development methodologies aiming to more nimble and lighter development processes, making them more responsive to change.

Agile authors built their methodologies on four principles. First, the main objective is to develop software that satisfies the customers, through continuous delivering of working software, and getting feedback from customers about it. The second principle is accepting changes in requirements at any development stage, so that customers would feel more comfortable with the development process. The third principle is the cooperation between the developers and the customers (business people) on a daily basis throughout the project development. The last principle is developing on a test-driven basis; that is to write test prior to writing code. A test suite is run on the application after any code change [18].

Agility in short means to strip away as much of the heaviness, commonly associated with traditional software development methodologies, as possible, in order to promote quick response to changing environments, changes in user requirements, accelerate project deadlines, and the like [16]. Agile methods have proved to be ten times more productive than traditional development models and this has been achieved in the co-located teams [17].

Probably the most used agile methods are Scrum and Extreme Programming, or XP. They can and many times are used together as Scrum is focused with project management techniques and XP is more focused on the actual development work [18].

Basically agile methodologies prefer software development over documentation. It believes in delivering many versions of the software in short iterations, then updates the software according to the customer's feedback. This methodology helps in overcoming the problems like frequent changes, faster development and user satisfaction.

## 4   COMPATIBILITY ISSUES OF DISTRIBUTED DEVELOPMENT (or GSD) WITH AGILE DEVELOPMENT

Both agile and distributed software developments (and GSD)  are growing trends as software business requires quicker quality production at a cheaper price. Distributed development is a fact of life for many agile teams. Most of the agile methodologies (e.g. scrum) assume that the team is located in a single room [6]. Unfortunately this principle does not fit in the real scenario where agile teams are also distributed across the globe. Factors like expanding business to new markets (Global markets through mergers and acquisitions), creating high quality employee pool, reduced costs through outsourcing to regions with cheaper development overheads are the main driving forces for organizations opting for distributed development.

In the 2008 State of Agile Development survey, conducted by VersionOne, 57% of respondents stated that their teams were distributed. Further 41% of respondents state that they were currently using or plan to combine agile with outsourced development. These facts clearly show that the current requirement of software industry is not in line with the agile concept of the entire agile team working in a single room [6]. Thus there is a need to extend the agile practices to globally distributed software development.

Agile methods can be beneficial when combined with distributed development. There are studies which show that agile principles help in overcoming some challenges faced by distributed development [7], [19], [20], [21], [22]. Agile when combined with DSD also brings some new challenges which will be discussed further.

### 4.1   Benefits achieved by combining agile with Distributed Development

Distribution of development (DSD/GSD) seems to cause decreased visibility of project status and agile process based on short continuous iterations make it easier to see the problems already on early stages of the project.

Continuous integration of software code, which is a central part of agile methods, also helps to reduce configuration management issues. Use of agile principles seems to have a positive effect on communication between teams as development in cycles makes it easier for participants to see the short term goals [23].

Sprint reviews can be an effective way to improve external communication while they help to share information about the feature and requirement dependencies between stakeholders.

Agile principles can even help create trust between different cultures involved in the process by constant communication and delivery of software [24]. According to a study made by Passivara, Durasiewicz and Lassenius quality of software and communication are improved and communication and collaboration is more frequent than before because of the Scrum methodology used in the project. Also the motivation of team members was reported to have increased [25].

Thus, agile in distributed teams has proved to be beneficial for project's quality and performance.

### 4.2 Challenges faced by agile distributed teams:

Along with the above mentioned benefits there are many new challenges that come when we combine agile with distributed development

Although there are business reasons for distributing a team, distribution leads to team dysfunction by inhibiting communication. Agile teams rely on intense person to person communication, both with team and the customer.

Thus, these two trends i.e. distributed development and agile approach face difficulty when it comes to compatibility issues.

**Documentation:** most offshore organizations favor the plan-driven approach where detailed requirements or designs are sent offshore to be constructed [7]. On the contrary agile teams tend downplay documentation from the observation that a large part of documentation effort is wasted. When teams are distributed, remote teams may miss out some details about the project and ultimately understanding suffers. Lack of rich conversation may make the information on the story cards insufficient for remote team members. Thus team members may need to supplement them with more details.

**Pair programming:** where two team members work side by side on the same code is another typical feature of agile practices. It is impossible to have this practice with distribute teams. Such teams will have to use some other equivalent practice.

**Different working hours:** Another major challenge comes for the teams which span over different time zones. There are times when one team member is available and other team member is not. Working hours of such teams need to be aligned which will help to increase clarity and avoid rework.

**Training on Agile Practices**: The impact of communication gap is felt more when the remote team members are new. They need to be trained on the agile practices like test driven development. Training is easy to give when teams are co-located. Distributed teams have to find alternatives to such trainings provided to the team members. Cultural differences amongst the team members further makes the situation difficult.

**Distribution of work:** One major challenge is that work distribution should not take place according to the location. This will lead to an architecture which will start reflecting the team's geographical distribution (Conway's law). Different locations will become overspecialized in a particular component. Ultimately it will become difficult to complete the user stories within an iteration as different parts of the team will have to complete specific pieces of work on a critical path to complete a user story. Thus, the teams need to distribute the work relating to single story across the whole team, regardless of geography, and think in terms of user stories not system components. This is the most challenging areas for distribute teams as they will have to work more closely across the geographical boundaries which will lead to reduced gaps in the functionality components that fall between components [6].

A corollary to this is to not divide teams by horizontally (having one team do presentation and another do database). In general a functional staff organization is preferred - and remote teams exacerbate the problems of dividing teams by layers. That means that it's important to separate distributed teams by modules that are as loosely coupled as possible [7].

## 5 TOOLS AND TECHNIQUES FOR AGILE DISTRIBUTE DEVELOPMENT

In order to create an effective distributed agile team, various communication barriers have to be overcome. Many distributed teams fail to work because they behave as if they are collocated and do not effectively address the additional communication burdens placed on them. Many of the core communication challenges faced require both commitment on the part of the team to improve and the support of additional practices and tools.

Following are various practices which can help to overcome the problems introduced by agile distributed software development methods.

**Improve Communication:** improved communication is the key to successful distributed teamwork. Minimize the overhead of setting up a meeting across the locations by having a conference phone and projector easily available. The whole team can get together easily fro impromptu meetings. Video conferencing if available is a better option than voice conferencing.

Team members can use hands free headsets, Web cams and application sharing softwares to allow them to work together remotely. Instant messenger can be used for synchronous communication and e-mail for asynchronous communication.

The way the team members communicate has to be changed. In collocated teams usually the communication is informal (verbal). This has to be changed to formal nonverbal communication. This approach is helpful in case of GSD as it helps in overcoming the problem of non overlapping working hours.

We can also have communication as an explicitly part of someone else's duties on the team. This could be done by assigning a representative for a remote subteam to help to catch up with the missed hallway communication.

Meeting formats may also need to change. The daily standup meetings may be overly long as the issues which are related to the whole team only are discussed. Other issues may be discussed after the meeting. The team members who are not involved the meeting can leave the room and get on to their work.

**Use Contact Visits:** Face-to-face communication can be done by project teams meeting on collocation for the first few iterations. This helps the team members to know each other and build trusts and rapport. This helps because many key decisions are taken at the beginning of the project. Personnel bonds can be refreshed by periodically bring the whole team back together during the course of project. Bringing the team together for the last couple of iterations helps in making the release and shipping of the final deliverable much smoother [6], [7].

**Team Distribution:** significant time zone differences introduce communication blackouts into the team's day during which the team in not available. A time zone distribution of three to four hours is workable with the whole team sharing either morning or afternoon hours. Team needs to make best use of overlapping times for meetings.

If the team is distributed offshore and the time zone distribution is large, then there could be a team representative who has worked with the remote team and has a good rapport with them. This representative will be the core of the team and will attend daily standup meetings. He needs to have active discussions and meetings in the team room and then pass the results to the offshore team. This work can lead to fatigue and burnouts.

An alternative for synchronized office hours has been suggested in an article by Hubert Smiths. He suggests using nested scrum with multilevel reporting and multiple daily scrum meetings [19]. The aim of all of these modifications is to help teams adjust to new methods and projects and help minimize misunderstandings between interest groups.

**Focus on Team coaching:** Distributed teams encounter more challenges and need more help to enable them to stick with the core practices employed by agile teams. Many distributed teams abandon key practices because they seem too hard. Having one person on the team who is committed to the coach role is vital in keeping a distributed team on the right path.

Many of the core Extreme Programming development practices serve to reinforce each other so it is important not to simply discard a practice, rather it should be modified or replaced with something equivalent. For example: pair programming might be augmented or replaced by code reviews and story cards may need complete information usually gained from conversations.

The coach's role is to remind the team of the underlying principles and help guide them to adapt their practices when the temptation might be to abandon them.

**Distribution of work:** One of the challenges when the team distributes work is not to do so according to location. The architecture will begin to reflect the team's geographical distribution (according to Conway's Law). Different locations will become over specialized in particular components.

Distributed teams should continue to think about their work in the context of completing user stories not adding features to components. They need to consciously distribute tasks relating to a single story across the whole team, regardless of geography, and think in terms of user stories not system components. Avoid breaking user stories into tasks and then assigning tasks according to geography and/or skill sets. Over time this would build up knowledge silos leaving the team with new work that can only be done by one or two people

**Documentation:** maintaining valuable documentation may also improve DSD team collaboration process while using agile practices [19], [26], [30], [31]. For example providing user stories with use case diagrams in globally accessible backlogs helps to reduce misunderstandings and improve team collaboration. Various tools like issue tracker (e.g. Jira), project management tool (e.g. Scrum works) also helps in maintaining documentation and good transparency [26].

**Use of tools:** Various tools have been suggested both formal and informal communication and project support.

Agile teams cannot rely on sticky notes on a task board or a burndown chart on the wall for project tracking if they are not in the same room. Similarly designs and diagrams need to be shared across multiple locations.

The decision to distribute a team needs to be coupled with the commitment to provide the team with the tools it needs to maximize communication and the expectation that it will take them some time to optimize around them.

The suggested tools can be categorized by their main function:

- Social networking tools: different social software tools and social networking tools enable group interactions which also include everything from email to video conferencing. These tools can be categorized by their main function:
- Communication tools: e-mails, instant messengers
- Software configuration management tools: repositories and version controlling tools.
- Bug and issue tracking databases: that contains the information about bugs found
- Knowledge centers: containing technical references and frequently asked questions.
- Collaborative development environments: providing project workspaces and standardized worksets e.g. project repositories and project management tools are recommended as necessary solutions in distributed agile projects.

According to Fowler, out of various ways to hold common information, wikis work well because they are simple to use, can be worked with any browser, and are simple to set up. Moreover wikis by nature unstructured, which is a benefit. The team can use their own structures to keep information as the project grows [7].

# 6 CONCLUSION

With distributed agile development it is possible to tap into new global markets and make best use of globally available talent, while potentially reducing costs.

The decision to distribute your project should be a conscious one and the decision maker(s) must understand that in doing so they reduce the project's likelihood of success, increase the delivery time, and reduce the team's performance and increase its dysfunction. The risk/reward tradeoff needs to be clearly understood before deciding to distribute your team(s).

There are many benefits of using agile methods with distributed software development. It helps in evaluating and measuring the progress of the project and problems of the project are more easily noticed at the early stage. It also handles the problems related to communication challenges in GSD such as difficulties in initiations and maintenance of communication.

There are new challenges introduces when agile is combined with distributed software development like communication misunderstandings and effect of different time zones. Informal communication which works well in collocated agile teams is not possible in distributed teams. This also leads to lack of trust amongst the team members. Thus, extra efforts are required on behalf of the team members to maintain effective communications.

Increase in the amount of documentation also helps in achieving better co-ordination requirement clarification with use-cases and user-stories [28]. Thus, flexibility and adjustment of project is recommended instead of strictly following agile principles [30].Efforts should be made bring the distributed team members together at the start, end and at other pivotal points during the project. This helps in building shared understanding of the problem domain and also working relationships within the team [6].

Agile teams work on user stories and not on component based tasks. Efforts should be made to distribute the work according to user stories and not by component. Agile development is hard and distributed development adds to the difficulty in software development. The team should have someone to coach them and guide them to move on the track of agile practices [6].

Providing the right kind of tools to the distributed team also helps. Some kind of online management tool is absolutely necessary for keeping up and tracking work on distributed teams [27].

Hence modified agile techniques if used consciously can be useful in overcoming the challenges faced by distributed software development (or GSD). The two software development methodologies i.e. agile and distributed development can be combined to give appreciable benefits to software industry in terms of increased production of high quality software by using optimized resources as compared to the traditional development models.

**Mrs. Suprika Vasudeva Shrivastava:** is M.E Computer Science and Engineering (I.T) (2004), B.E in Computer Science and Engineering (2001), Diploma in Computer programming and Application (1998). She has been into academics since past 8 years. Worked as lecturer in Shri Guru Gobind Singh College, Chandigarh (2001), Dehradun Institute of Engineering and Technology (D.I.T), Dehradun (2002,2004). She is currently working as Assistant Professor in Symbiosis Centre for Information Technology (SCIT) since 2005. She has total 8 papers (research and review) papers published in various conferences and journals of national and International level. One of her publication is available on IEEE Explore. She recently participated in a sponsored research project on project management for PMI, India. She is also working as a reviewer for SCIT journal. Her current areas of research are Risk Management, Agile Software Development and Software Quality.

**Dr. (Mrs.) Hema Date:** is a PhD. (National Institute of Industrial Engineering (NITIE), Mumbai), PGDIE (NITIE), and B.E

(Mumbai). She is currently working as Associate Professor in the Department of Information Technology, NITIE, Mumbai. She has more than 30 publications at National and International level. She has done various sponsored research projects for Ministry for Human resource and Development, India. Her research areas are Software Engineering, MIS, Knowledge Management, Expert Systems. She is a member of Indian Institute of Plant Engineering (IT Curriculum Development).